\documentclass[11pt,a4paper]{article}
\usepackage{graphicx}
\begin{document}
\title{\textsc{ 
Monopoly Market with Externality: an Analysis with Statistical Physics and Agent Based Computational Economics}}
\author {Jean-Pierre Nadal (1), Denis Phan (2) \\ Mirta B. Gordon(3)
and Jean Vannimenus (1)}

\vspace{2cm}
\date{ \scriptsize{
(1) Laboratoire de Physique Statistique, Ecole Normale Sup\'erieure, \\
24 rue Lhomond, 75231 Paris cedex 05, France\\
(nadal@lps.ens.fr, http//www.lps.ens.fr/\~{ }nadal)\\
(2) ENST Bretagne, ICI-UBO, Brest, France\\
(denis.phan@enst-bretagne.fr,  http//www-eco.enst-bretagne.fr/\~{ }phan)\\
(3) Laboratoire Leibniz-IMAG, \\
46, Ave. F\'elix Viallet, 38031 Grenoble Cedex 1, France\\
(mirta.gordon@imag.fr, web: http//www-leibniz.imag.fr/Apprentissage)\\
http//www-eco.enst-bretagne.fr/\~{ }phan/npgweiha2003.pdf \\
Presented at the 8th Annual Workshop on Economics with Heterogeneous Interacting Agents \\
(WEHIA 2003, May 29-31, Kiel)
}} 
\maketitle 
\thispagestyle{empty}

\begin{abstract}
\scriptsize{
We explore the effects of social influence in a simple market model
in which a large number of agents face a binary choice:
'to buy/not to buy' a single unit of a product at a price posted by a single
seller (monopoly market). We consider the case of {\em positive externalities}:
an agent is more willing to buy 
if the other agents with whom he/she interacts make the same decision.

We compare two special cases known in the economics literature as the
Thurstone and the McFadden approaches. We show that they correspond
to modeling the heterogenity in individual decision rules with,
respectively, annealed and quenched disorder. More precisely the first
case leads to a standard Ising model at finite temperature in a uniform
external field, and the second case to a random field Ising model (RFIM)
at zero temperature.

We illustrate some dynamic properties of these models, making use of
numerical simulations in an ACE (Agent based Computational Economics) approach, and 
we study analytically the
equilibrium properties in the mean field limit.

Considering the optimisation of profit by the
seller within the McFadden/RFIM model in the mean field limit, we
exhibit a new first order phase transition: 
if the social influence is strong enough, there is a regime where,
if the mean willingness to pay increases, or if the production costs decrease,
the optimal solution for the seller jumps from one with a high price and a
small number of buyers, to another one with a low price and a large number of buyers.

{\em This paper has been presented at the interdisciplinary conference
WEHIA 2003: it is intended to be  accessible to
economists and physicists - without, hopefully, discouraging any of them!}
}
\end{abstract}

\section{Introduction}
\label{sec.monopoly}

Following Kirman~\cite{Kirman97a,Kirman97b}, we view a market 
as a complex interactive system with a communication network. We consider 
a market with discrete choices~\cite{ADT}, and explore the effects of 
localised externalities (social influence) on its properties.  
We focus on the simplest case: a single homogeneous product and a single 
seller (monopoly). On the demand side, customers are assumed to be myopic 
and non strategic. The only cognitive agent in the process is the monopolist, 
who determines the price in order to optimise his profit. We obtain 
interesting and complex phenomena, that arise due to the interactions 
between heterogeneous agents. 

Agents are assumed to have idiosyncratic willingness-to-pay (IWP) which are described 
by means of random variables. We consider two different models: on one hand, the IWP
are randomly chosen and remain fixed, on the other hand, the IWP present independent temporal 
fluctuations around a fixed (homogeneous) value. 
We will show that they correspond to statistical physics models, in the former case with {\it quenched disorder}, 
and in the latter with  {\it annealed disorder}. 
In both cases we assume that the heterogeneous 
preferences of the agents are drawn from the same (logistic) distribution. 
The equilibrium states of the two models generally differ, except in the 
special case of homogeneous interactions with complete connectivity. In 
this special situation, which corresponds to the {\it mean-field approximation}
in physics, the expected aggregate steady-state is the same in both models. 
 
In the following, we first present the demand side, and then consider the 
optimisation problem left to the monopolist, who is assumed to know the 
demand model and the distribution of the IWP over the population, but cannot observe the individual (private) values.

\section{Simple models of discrete choice with social influence}
\label{sec.models}

We consider a set $\Omega_N$ of $N$ agents with a classical linear IWP 
function~\cite{PhanPajotNadal03}. Each agent $i\in \Omega_N$ either buys 
($\omega_{i}=1$) or does not buy ($\omega_{i}=0$) one unit of the 
single given good of the market. A rational agent chooses $\omega_{i}$ 
in order to maximise his {\it surplus function} $V_{i}$:
\begin{equation}
\label{eq.surplus}
\max_{\omega_{i}\in \{0{,1}\}}V_{i}=\max_{\omega_{i}\in \{0{,1}\}}\omega_{i}
(H_i+ \sum_{k \in \vartheta_{i}} J_{ik} \omega_{k}-P),  
\end{equation}
where $P$ is the price of one unit and $H_{i}$ represents the idiosyncratic 
preference component. Some other agents $k$ within a subset $\vartheta_{i}\subset\Omega_N$  
(such that $k \in \vartheta_{i}$) hereafter called neighbours of $i$, influence agent $i$'s preferences 
through their own choices $\omega_k$. This social influence is represented here by a weighted sum 
of these choices. Let us denote $J_{ik}$ the corresponding weight {\it i.e.} the marginal social influence of the decision of agent $k \in \vartheta_{i}$ on agent $i$.
When this social influence is assumed to be positive ($J_{ik}>0$), it is possible, following Durlauf~\cite{Durlauf97},
to identify this external effect as a {\it strategic complementarity} in agents' choices~\cite{BGK}. 

For simplicity we consider here only the case of {\it homogeneous} influences, 
that is, identical positive weights $J_{ik}=J_\vartheta$ and 
identical neighbourhood structures $\vartheta$ of cardinal $n$, for all the 
agents. That is,
\begin{equation}
\label{eq.Jtheta}
J_{ik}=J_\vartheta \equiv J/n > 0 \,\,\,\,\,\,\, \forall i\in \Omega_N, \,\,\, k \in \vartheta_i
\end{equation}

\subsection{Psychological {\em versus} economic points of view}

Depending on the nature of the idiosyncratic term $H_i$, the discrete choice 
model (\ref{eq.surplus}) may represent two quite different situations. Following 
the typology proposed by Anderson {\it et al.}~\cite{ADT}, we distinguish a 
``psychological'' and an ``economic'' approach to individual choices. Within the 
psychological perspective (Thurstone~\cite{Thurstone}), the utility has a {\it stochastic}
aspect because ``there are some qualitative fluctuations 
from one occasion to the next... for a given stimulus'' (this point of view 
will be referred to hereafter as the {\it TP-model}). On the contrary, for 
McFadden~\cite{McFadden} each agent has a willingness to pay that is 
{\it invariable} in time, but may differ from one agent to the other. 
In a ``risky'' situation the seller cannot observe each specific 
IWP, but knows its statistical distribution 
over the population (we call this perspective the {\it McF-model}). Accordingly, 
the TP and the McF perspectives 
differ in the nature of the individual willingness to pay. 

In the TP model, the idiosyncratic preference has two sub-components:
a constant deterministic term $H$ (the same for all the agents), and a time- and 
agent-dependent additive term $\epsilon_{i}(t)$, so that $H_i=H+\epsilon_i$.
The $\epsilon_{i}(t)$ are i.i.d. random variables of zero mean; 
in the simulations they are refreshed at each time step (asynchronous updating). 
Agent $i$ decides {\it to buy} according to the conditional probability 
\begin{equation}
\label{eq.pcond}
P(\omega_i=1 | z_i(P,H))={\mathcal P}(\epsilon_i > z_i(P,H))=1-F(z_i(P,H)),
\end{equation}
with
\begin{equation}
\label{eq.z_i}
z_i(P,H) = P-H- J_\vartheta \; \sum_{k \in \vartheta_i} \omega_k,
\end{equation}
where $F(z_i)={\mathcal P}(\epsilon_i \leq z_i)$ is the cumulative distribution of the random variables 
$\epsilon_i$. In the 
standard TP model, the agents make repeated choices, and the time varying components 
$\epsilon_{i}(t)$ are drawn at each time $t$ from a logistic distribution with zero mean, and variance 
$\sigma^2=\pi^2/(3 \beta^2)$: 
\begin{equation}
\label{eq.logistic}
F(z) = \frac{1}{1+ \exp(-\beta \; z)}.  
\end{equation}

In the McF model, the private idiosyncratic terms $H_{i}$ are randomly 
distributed over the agents, but remain fixed during the period under
consideration. There are no temporal variations: the $\epsilon_{i}$ are 
strictly zero. In analogy with the TP model, it is useful to introduce 
the following notation: $H_{i}=H+\theta_{i}$, and to assume that the 
$\theta_{i}$ are logistically distributed with zero mean and variance 
$\sigma^2=\pi^2/(3 \beta^2)$ over the population. This assumption implies:
\begin{equation} 
\lim_{N \to \infty } \frac{1}{N} \sum_i \theta _i = 0 \;\;{\rm and}  \;\; \lim_{N \to \infty} \frac{1}{N} \sum_i H_i = H.
\end{equation}
Thus, the correspondence between models is better the larger the number of agents, and is strict only in the limit of an infinite population. In practice, the simulations presented in the following sections show that the theoretical predictions are already verified for population sizes of the order of some hundreds.

For a given distribution of choices in the neighbourhood $\vartheta_i$, and for a given price,
the customer's behaviour in the McF model is deterministic. An agent buys if:
\begin{equation}
\label{eq.theta_i}
\theta_i > P-H- J_\vartheta \; \sum_{k \in \vartheta_i} \omega_k.
\end{equation}

\subsection{`Annealed' {\em versus} `quenched' disorder}
Since we have assumed isotropic (hence symmetric) interactions,
there is a strong relation between these models and Ising type models
in Statistical Mechanics, which is made explicit if we change
the variables $\omega_{i} \in \left\{ 0,1\right\}$ into
{\it spin} variables $s_{i} \in \{ \pm 1\}$ through
$ \omega_{i}=\frac{1+s_{i}}{2}$. 
All the expressions in the present paper can be put in terms of either
$s_{i}$ or $\omega_{i}$ using this transformation.
In the following we keep the encoding $\omega_i \in \left\{ 0,1\right\}$. 
The assumption of {\it strategic complementarity} in economics corresponds to
having ferromagnetic couplings in physics (that is, the interaction $J$ between 
Ising spins is positive). In this case, the spins $s_i$ (and consequently the choices 
$\omega_i$) all tend to take the same value. 

In physics, the TP model corresponds to a case of {\it annealed disorder}. 
Having a time varying random idosyncratic component is equivalent to introducing a
stochastic dynamics for the Ising spins. In the particular case where $F(z)$ in (\ref{eq.pcond}) 
is the logistic distribution, we obtain an Ising model 
in a uniform (non random) external field $H-P$, at temperature $T=1/\beta$. 
The McF model has fixed heterogeneity; it is analogous to a {\it Random 
Field Ising Model} (RFIM) at zero temperature, that is, with deterministic 
dynamics. The RFIM belongs to the class of {\it quenched disorder} models: the 
values $H_{i}$ are equivalent to random time-independent local fields. 
The ``agreement'' among agents favored by the ferromagnetic couplings may 
be broken by the influence of these heterogeneous external fields $H_{i}$. Due to 
the random distribution of $H_{i}$ over the network of agents, the resulting 
organisation may be complex. Thus, from the physicist's point of view, the TP and 
the McF models are quite different: uniform field and finite temperature in the 
former, random field and zero temperature in the latter. The properties of 
disordered systems have been and still are the subject of numerous studies 
in statistical physics. They show that annealed and quenched disorder can 
lead to very different behaviours. 

The TP model is well understood. In the case where the agents are situated on the 
vertices of a 2-dimensional square lattice, and have four neighbours each, there 
is an exact analysis of the model for $P=P_n$ (the unbiased case, (\ref{eq.Pn}), see below)
due to Onsager~\cite{Onsager}. Even if an 
analytical solution of the optimization problem (\ref{eq.surplus}) for an 
arbitrary neighbourhood does not exist, the {\it mean field} analysis gives 
approximate results that become exact in the limiting situation where every 
agent is a neighbour of every other agent ({\it i.e.} all the agents are interconnected 
through weights (\ref{eq.Jtheta})). On the contrary, the properties of the McF model 
are not yet fully understood. A number of important results have been published 
in the physics literature since the first studies of the RFIM by Aharony and 
Galam~\cite{GalamAharony80,GalamAharony81} (see also \cite{Galam82}, 
\cite{Sethna93}). 
Several variants of the RFIM have already been used in the context of 
socio-economic modelling~\cite{GGF,Orlean,Bouchaud,WeSt}. 

\subsection{Static {\em versus} dynamic points of view}
Hereafter, we restrict our investigation to the McF model in the case of a ``global''
externality. That is, we consider {\it homogeneous interactions} and {\it full connectivity}, 
which is equivalent to the {\it mean field} model in physics. Within this general framework,
we are interested in two different perspectives. 
First we consider a static point of view: solving for the equality between
demand and supply we determine the set of possible economic equilibria.
This will allow us to analyse in section \ref{sec.supply_side} the optimal strategy of the monopolist,
as a function of the model parameters.

Next (section \ref{sec.dynamic}) we consider the market's dynamics assuming 
myopic and non strategic agents:
based on the observation of the behaviour of the other agents at time $t-1$, 
each agent decides at time $t$ to buy or not to buy. 
We show that, in general, the market
converges towards the static equilibria of the preceding section,
except for a precise range of the parameter values where 
interesting static as well as dynamic features are observed.

These two kinds of analysis correspond in Physics to the study of
the thermal equilibrium properties within the {\it statistical ensemble}
framework on the one hand, and the out of equilibrium dynamics
(which, in most cases, approaches the static equilibrium through a relaxation process)
on the other hand, respectively.

\section{Aggregate demand}
\label{sec.demand}
As discussed in the preceding section, we consider the full connectivity case with isotropic interactions given by (\ref{eq.Jtheta}) with cardinal $n=N-1$, in the limit of a very large number of agents. In this limit the 
{\it penetration rate} $\eta$, defined as the fraction of customers that choose to buy at a given price: 
$\eta \equiv lim_{N \rightarrow \infty} \sum_{k=1}^{N} \omega_k /N$, 
can be approximated by $\eta \approx \sum_{k\in \vartheta} \omega_k /(N-1)$, which is proportional to the social influence term of the agents' surplus function (eq. (\ref{eq.surplus})). Equation (\ref{eq.theta_i}) may thus be replaced by 
\begin{equation}
\label{eq.theta_i2}
\theta_i > P-H- J\; \eta.
\end{equation}
For the following discussion it is convenient to identify the 
{\it marginal customer}, indifferent between buying and not buying. 
Let $H_{m}=H+\theta_{m}$ be his IWP. 
This {\em marginal customer} has zero surplus ($V_{m}=0$), that is: 
\begin{equation}
\label{eq.thetam}
\theta_m = P-H-J \; \eta \equiv z.
\end{equation}
Thus, an agent $i$ buys if $\theta_i > \theta_m$, and does not buy otherwise. 
Equations (\ref{eq.theta_i2}) and (\ref{eq.thetam})
allow us to obtain $\eta$ as a fixed point:
\begin{equation}
\label{eq.fp}
\eta =1-F(z)
\end{equation}
where 
$z$, defined in (\ref{eq.thetam}) depends on $P$, $H$, and $\eta$, and $F(z)$ is the cumulative distribution of the IWP around the average value $H$. 
Note that this (macroscopic) equation is formally equivalent to the (microscopic)
individual expectation that $\omega_{i}=1$ in the TP case (\ref{eq.pcond}). 
Using the logistic distribution for $\theta_i$, we have:
\begin{equation}
\label{eq.fp2}
\eta =\frac{1}{1+\exp\,(+\,\beta \; z)}
\end{equation}

Equation (\ref{eq.fp}) allows us to define $\eta$ as an implicit function of the price through
\begin{equation}
\label{eq.implicit}
\Phi(\eta,P) \equiv \eta(P)+ F(P-H-J \; \eta(P)) - 1 =0.
\end{equation}
The shape of this (implicit) demand curve can be evaluated 
using the implicit derivative theorem:
\begin{equation}
\label{eq.implicitDerivative}
\frac{d\eta(P)}{dP}=\frac{-\partial\Phi/\partial P}{\partial\Phi/\partial\eta}=\frac{-f(z)}{1-J\, f(z)}
\end{equation}
where $f(z) = dF(z)/dz$ is the probability density.

Since for a given $P$, equation (\ref{eq.fp2}) defines
the penetration rate $\eta$ as a fixed-point, inversion of this equation gives an {\it inverse demand function}:
\begin{equation}
\label{eq.price_d}
P^d(\eta) = H+J \; \eta + \frac{1}{\beta} \ln \frac{1-\eta}{\eta} 
\end{equation}

At given values of $\beta$, $J$ and $H$, for most values of $P$, (\ref{eq.fp2})
has a unique solution $\eta(P)$. 
However for $\beta J > \beta J_B \equiv 4$, there is a range of prices
\begin{equation}
\label{domain}
P_{1}(\beta J, \beta H) < P < P_{2}(\beta J, \beta H)
\end{equation}
such that, for any $P$ in this interval, (\ref{eq.fp2})
has two stable solutions and an unstable one. 
The limiting values $P_1$ and $P_{2}$ are the particular price values
obtained from the condition that eq. (\ref{eq.fp2})
has one degenerate solution:
$$\eta = 1 - F(z), \;\mbox{and} \;\; \frac{d (1-F(z))}{d \eta} = 1.$$
This gives
$\beta J \eta ( 1 - \eta) = 1$
(together with eq. (\ref{eq.fp2})).
This equation has two solutions, $\eta_{2} \; \leq 1/2 \leq \; \eta_{1}$,
\begin{equation}
\label{eta12}
\eta_{i} = \frac{1}{2} \left[1 \pm  \sqrt{1- \frac{4}{\beta J}}\right]\; ; \;\;i \in \;\{1,2\} 
\end{equation}
Note that $\eta_{i}$ depends only on $\beta J$.
Then, the limiting prices $P_{i}$ are equal to the inverse demands associated with these
values $\eta_{i}$, which are, from (\ref{eq.price_d}): 
\begin{equation}
\label{P12}
P_{i} = H + J \eta_{i} + \frac{1}{\beta} \; \ln[\frac{1-\eta_{i} }{\eta_{i} }]\; ; \;\;i \in \;\{1,2\} .
\end{equation}
Note that these limiting prices are not necessarily positive.

It is interesting to note that the set of equilibria is the same as
what would be obtained if agents had rational expectations about the choices of the others:
if every agent had knowledge of the distribution of the $H_i$, 
he could compute the equilibrium state
compatible with the maximisation of his own surplus, 
taking into account that every agent does the same,
and make his decision (to buy/not to buy) accordingly. 
For $\beta J < \beta J_B $ every agent could thus anticipate the value of $\eta$
to be realized at the price $P$, and make his choice according to (\ref{eq.theta_i2}). 
For $\beta J > \beta J_B $, however, if the price is set within the interval $[P_1, P_2]$, 
the agents are unable to anticipate which equilibrium
will be realized, even though the one with the largest value of $\eta$
should be preferred by every one (it is the Pareto dominant equilibrium).

\section{Supply side}
\label{sec.supply_side}
On the supply side, we consider a monopolist facing 
heterogeneous customers in a risky situation where the monopolist has perfect 
knowledge of the functional form of the agents' surplus functions and the 
maximisation behaviour (\ref{eq.surplus}). He also knows the statistical (logistic) 
distribution of the idiosyncratic part of the reservation prices 
($H_{i}$). In the special case of global externality considered here, where the
interactions are the same for all customers, as in equation (\ref{eq.Jtheta}) with $n=N-1$, as just seen,
the TP model and the McF one have the same equilibrium states.
Indeed, the monopolist cannot observe any {\it individual} reservation
price. He observes only the cumulated result of the individual choices (which are either to buy or not to buy).
As a result, the conditional probability for an agent taken at random by
the monopolist to be a customer is formally equivalent in both cases.
Accordingly, hereafter we limit ourselves to the case where the demand 
follows the McF model, in the limit of full connectivity.
The social influence on each individual decision is then close 
to $J \; \eta$, and the fraction of customers $\eta$ is 
observed by the monopolist. That is, for a given price, the expected 
number of buyers is given by equation (\ref{eq.fp}). 

\subsection{Profit maximisation}
Let $C$ be the monopolist cost for each unit sold, so that 
\begin{equation}
p \equiv P-C 
\end{equation}
is his profit {\it per unit}. 
Since $P-H=(P-C)-(H-C)$, defining 
\begin{equation}
h\equiv H-C,
\end{equation}
we can rewrite $z$ in (\ref{eq.z_i}) as:
\begin{equation}
\label{eq.z}
z = p\; -\; h\; -\; J\; \eta.
\end{equation}
Hereafter we write all the equations in terms of $p$ and $h$.

Since each customer buys a single unit of the good, 
the monopolist's  total expected profit is $p \; N \; \eta$. Thus, in this mean 
field case, the monopolist's profit is proportional to the total number of 
customers. He is left with the following maximisation problem: 
\begin{equation}
p_M = \arg \max_{p} \Pi(p), 
\end{equation}
where $N\; \Pi(p)$ is the expected profit, with: 
\begin{equation}
\Pi(p) \equiv p \; \eta(p),
\end{equation}
and ${\eta (p)}$ is the solution to the implicit equation (\ref{eq.fp}). 
If there is no discontinuity in the demand curve $\eta(p)$ (hence for $\beta J \leq 4$),
$p_M$ satisfies 
$d\Pi(p)/dp = 0$, 
which gives $d\eta/dp=-\eta/p$ at $p=p_M$. Using the
implicit derivative (\ref{eq.implicitDerivative}), we obtain at $p=p_M$: 
\begin{equation}
\label{ImplicitDeriv}
\frac{f(z)}{1-J f(z)}=\frac{\eta}{p},
\end{equation}
where $z$, defined in (\ref{eq.z}), has to be taken at $p=p_M$.

Because the monopolist observes the demand level $\eta$, 
we can use equation (\ref{eq.fp}) to replace $1-F(z)$ by $\eta$.
After some manipulations, equation (\ref{ImplicitDeriv}) gives an inverse supply function $p^s(\eta)$:
\begin{equation}
\label{eq.price_s}
p^s(\eta) = \frac{1}{\beta \; (1-\eta)} - J \; \eta   
\end{equation}
We obtain $p_M$ and $\eta_M$ as the intersection between supply (\ref{eq.price_s}) and demand (\ref{eq.price_d}): 
\begin{equation}
\label{eq.price}
p_M = p^s(\eta_M) = p^d(\eta_M),
\end{equation}
where $p^d \equiv P^d-C$.

The (possibly local) maxima of the profit are the solutions of (\ref{eq.price}) for which
\begin{equation}
\frac{d^2 \Pi}{dp^2} < 0.
\label{eq.stab}
\end{equation}
It is straightforward to get the expression
for the second derivative of the profit:
\begin{equation}
\frac{d^2 \Pi}{dp^2} = - 2 \frac{\eta}{p} \left[ 1 + \frac{2 \eta - 1}{2 \beta p (1 - \eta)^2} \right],
\label{d2pi_L}
\end{equation}
from which it is clear 
that the solutions with $\eta > 1/2$ are local maxima. For
$\eta < 1/2$, condition (\ref{eq.stab}) reads
\begin{equation}
\frac{1 - 2 \eta}{2 \beta p (1 - \eta)^2} < 1.
\label{cond1}
\end{equation}
Making use of the above equations, this can also be rewritten as
\begin{equation}
2 \beta J \eta ( 1 - \eta)^2 < 1.
\label{cond2}
\end{equation}

For $\beta J > \beta J_B = 4$, the monopolist has to find $p=p_M$ which realises the programme:
\begin{equation}
\label{maxprofit}
p_M : \max \{\Pi_{-}(p_{-}^M), \Pi_{+}(p_{+}^M) \}
\end{equation}
\begin{equation}
\label{maxprofit+}
p_{+}^M = arg \max_{p} \Pi_{+}(p) \equiv p\;  \eta_{+}(p),
\end{equation}
\begin{equation}
\label{maxprofit-}
p_{-}^M = arg \max_{p} \Pi_{-}(p) \equiv  p \; \eta_{-}(p)
\end{equation}
where the subscripts $+$ and $-$ refer to the solutions of (\ref{eq.fp2})
with a fraction of buyers larger, respectively smaller, than $1/2$.

\begin{figure}[htbp]
\vspace{-1ex}
\centering
\begin{tabular}{|c|c|}
\hline
\includegraphics[width=6.2cm]{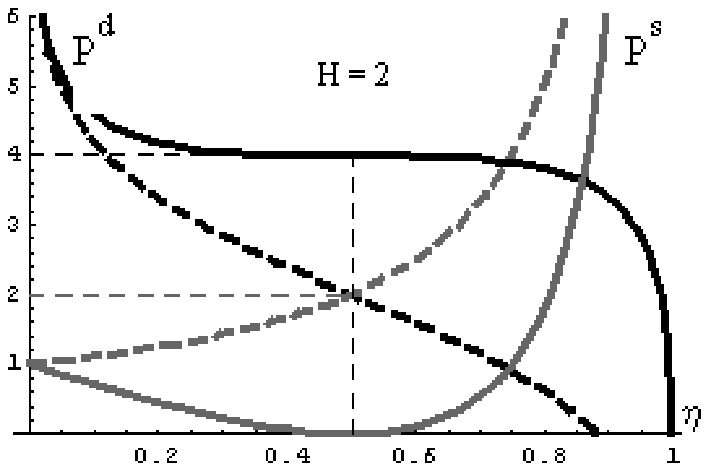}&
\includegraphics[width=6.2cm]{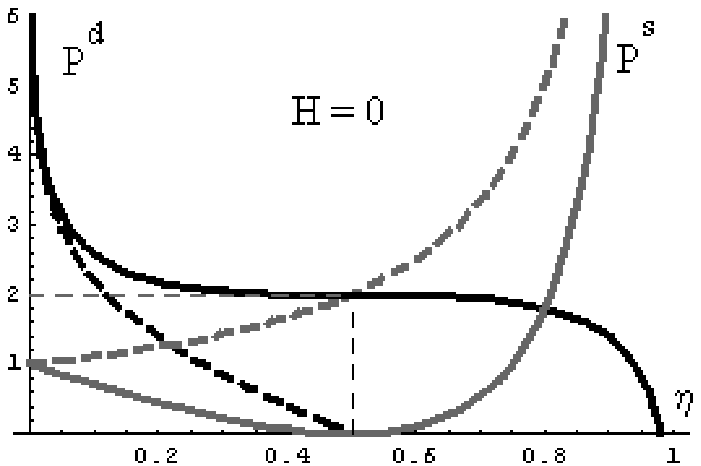}\\
\hline
\includegraphics[width=6.2cm]{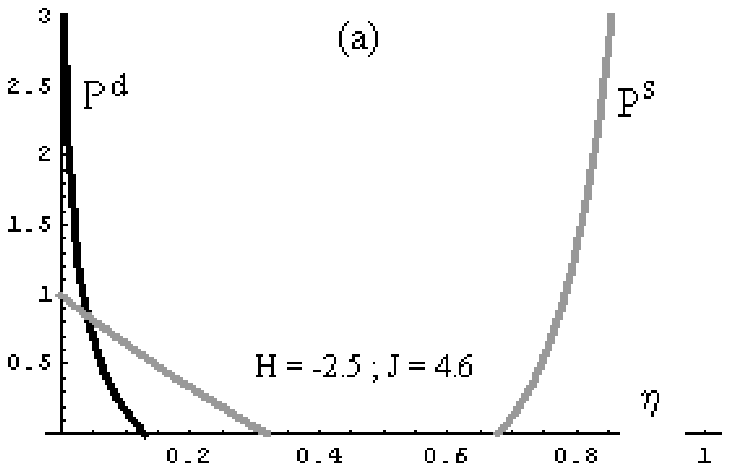}&
\includegraphics[width=6.2cm]{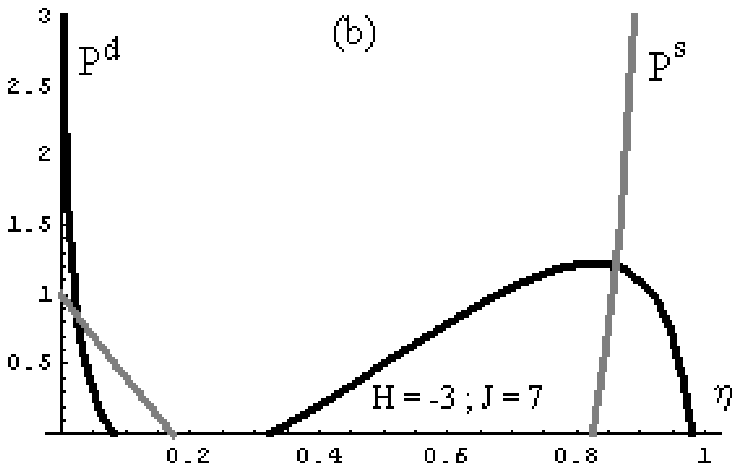}\\
\hline
\includegraphics[width=6.2cm]{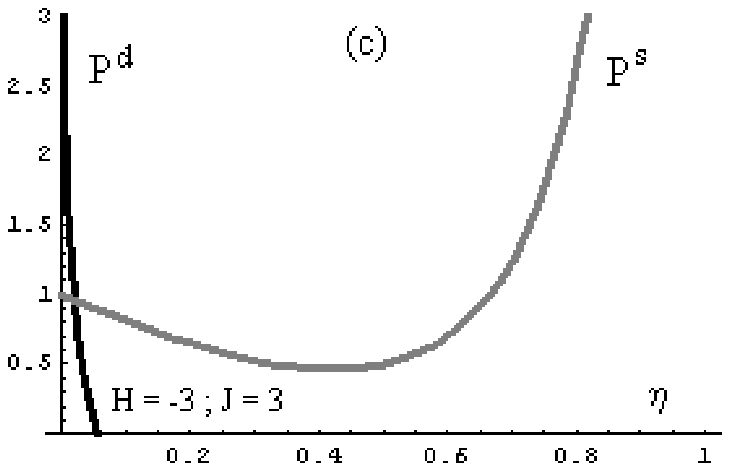}&
\includegraphics[width=6.2cm]{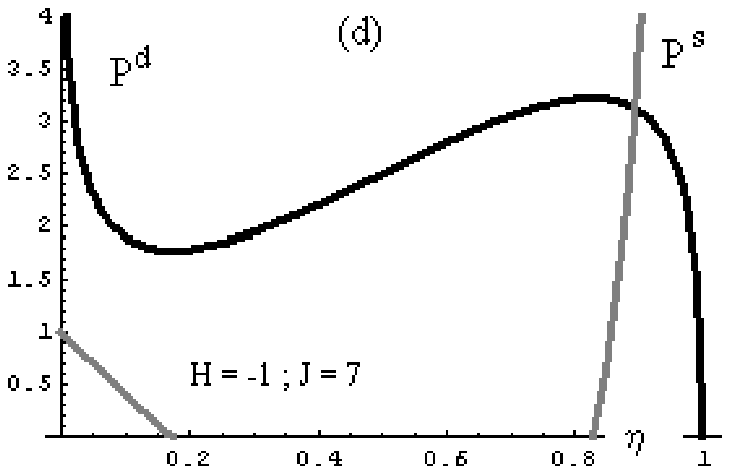}\\
\hline
\end{tabular}

\caption[]{ \scriptsize{Inverse supply and demand curves $p^s(\eta)$ and $p^d(\eta)$,
for different values of $h$ and $J$ ($\beta =1$, $C=0$ hence $h=H$). The equilibrium prices are obtained at the intersection between the demand (black) and the supply (grey) curves. \\
The two graphics on top illustrate the difference between a complete absence of externality 
($J=0$, dashed lines) and a strong externality ($J =4$, solid lines). 
The case $h=2$ (left) corresponds to a 
strong positive average of the population's IWP ($h=2$), whereas the population is neutral for 
$h=0$ (right).\\
The four graphics labelled (a) to (d) were obtained for values of $h$ and $J$ corresponding to the points (a) to (d) in the phase diagram (figure \protect{\ref{phasediagram.eps}}). They all have negative values of the average of the population's IWP ($h < 0$), so that in the absence of externality only few consumers would be interested in the single commodity.\\
(a) corresponds to the {\it coexistence region} between two local market equilibria 
in figure \protect{\ref{phasediagram.eps}}; 
but one of them (not shown) is not relevant since it corresponds to a negative price solution. 
(b) lies in the {\it coexistence region}; in this case, the optimal market equilibrium  is the one with high $\eta$.
(c) lies in the region with only one market equilibrium, with few buyers (small $\eta$).
(d) corresponds to a large social effect; the single market equilibrium has large $\eta$ and a high price.}}
\label{fig.market_configs}
\end{figure}

To illustrate the behaviour of these equations, figure \ref{fig.market_configs} 
represents several examples of inverse supply and demand curves corresponding to different market configurations. 

\subsection{Phase transition in the monopolist's strategy}

\begin{figure}
\centering
\includegraphics[width=0.65\textwidth]{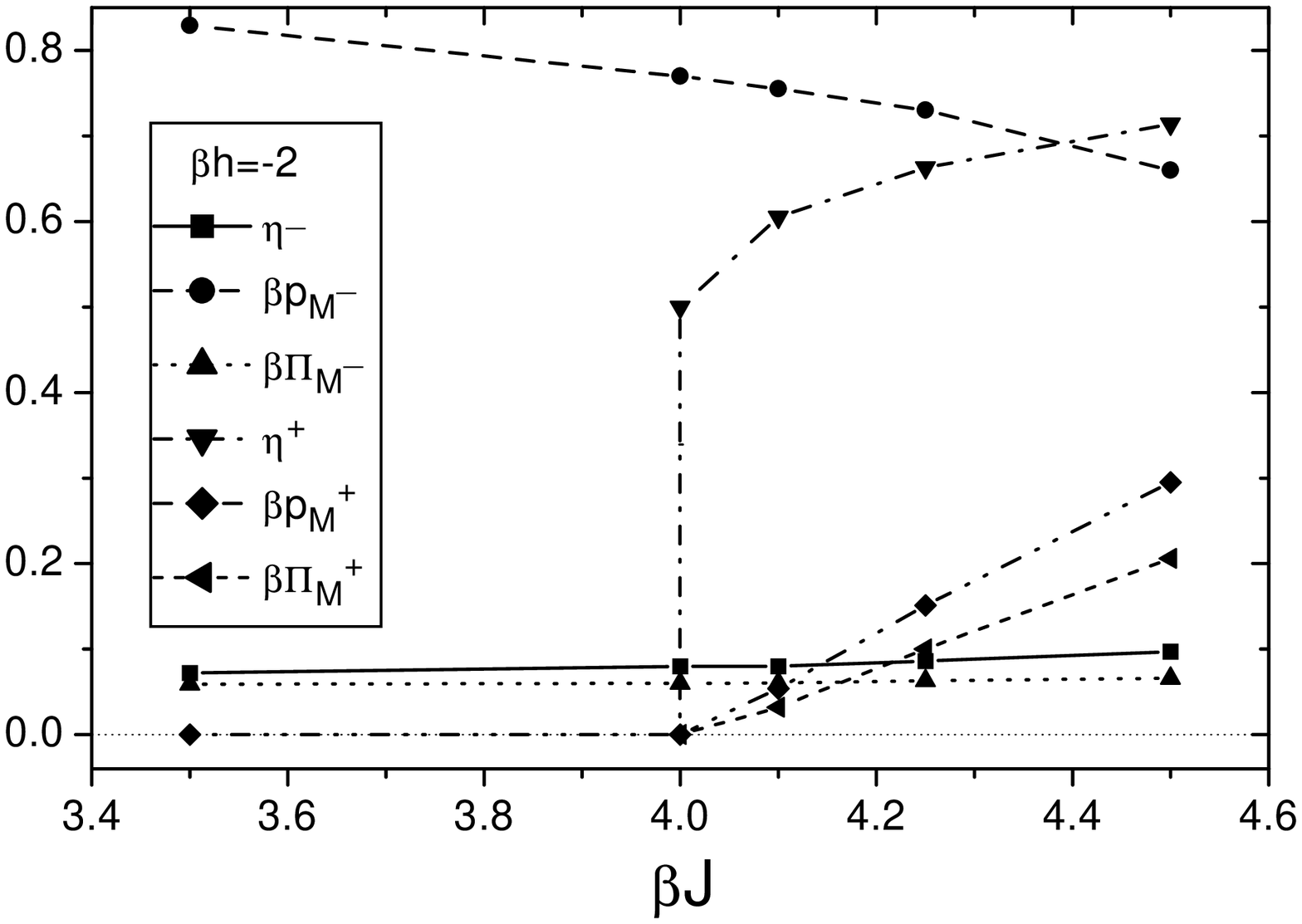}
\caption[]{ \scriptsize{ Fraction of buyers $\eta$, optimal price $\beta p_M$ and monopolist profit $\Pi_M$, as a 
function of the social influence, for $\beta h=-2$. The superscripts $-$ and $+$ refer to 
the two solutions of equations (\ref{eq.price}) that are relative maxima.}}
\label{fig.supply}
\end{figure}

\begin{figure}
\centering
\includegraphics[width=0.65\textwidth]{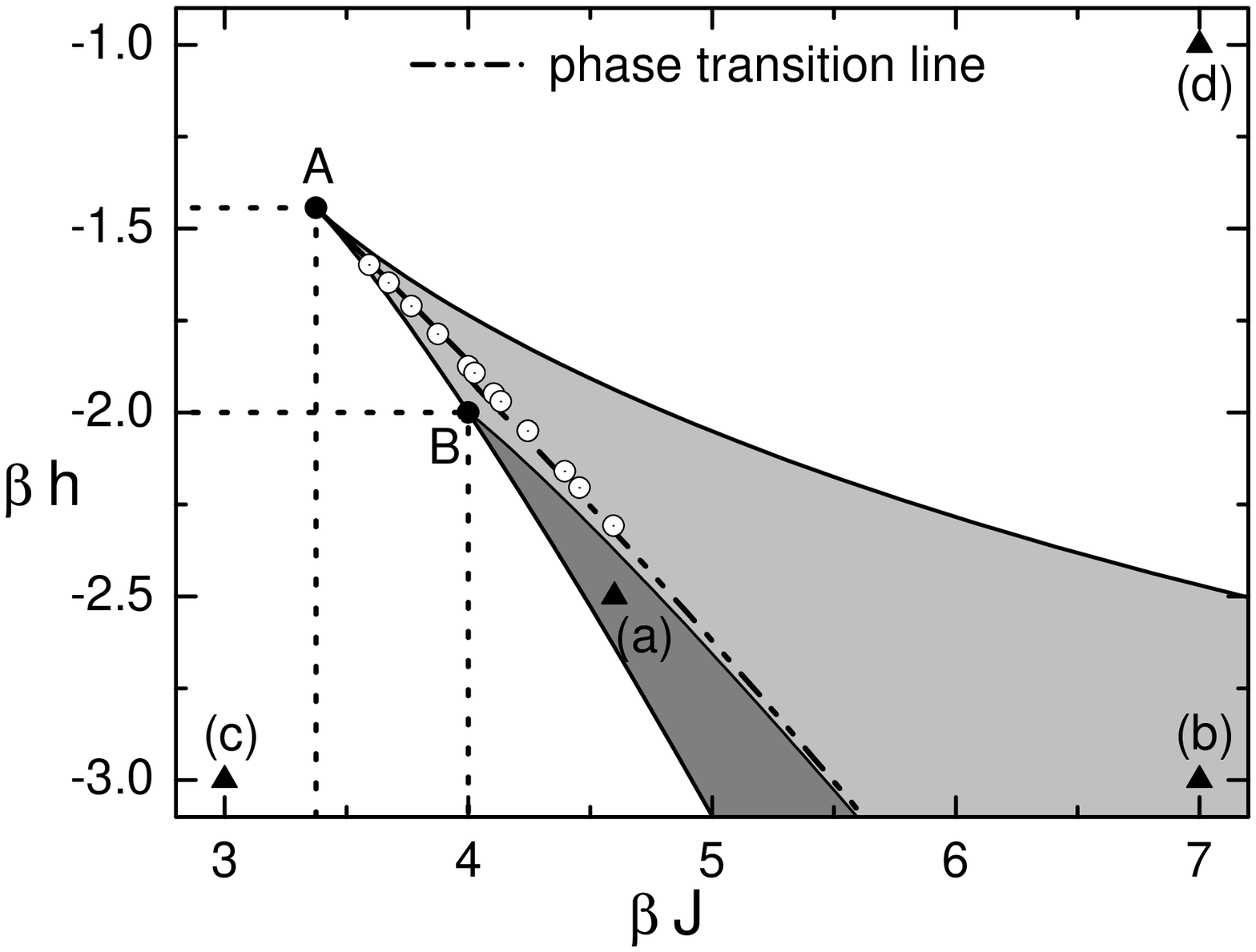}
\caption[]{ \scriptsize{
Phase diagram in the plane $\{\beta J, \beta h\}$: 
the grey region represents the domain in the parameter space where coexist two maxima of the monopolist's profit, a global one
(the optimal solution) and a local one. Inside this domain, as $\beta J$ and/or $\beta h$ increase, there
is a (first order) transition where the monopolist's optimum jumps from a high price, low penetration rate solution 
$\eta=\eta_-$ to one with low price, large $\eta=\eta_+$.
The circles on the transition line have been
obtained numerically, the smooth curves are obtained analytically (see 
the Appendix and \cite{extended} for details).
The points (a) to (d) correspond to the inverse supply and demand curves represented
in figure \protect{\ref{fig.market_configs}}.
In the white region, for $\beta J < 27/8$, the fraction of buyers, $\eta$,
increases continuously from $0$ to $1$ as $\beta h$ increases from $- \infty$ to $+ \infty$ (c-d).
At the singular point $A$, ($\beta J = 27/8$, $\beta h = -3/4 - \log(2)$), 
$\eta_+ = \eta_- = 1/3$. 
At point $B$ ($\beta J = 4$,  $\beta h = -2$), 
the local maximum with $\eta$ large appears with a null profit
and $\eta_+=1/2$.
In the dark-grey region below $B$, this local maximum 
exists with a negative profit, being thus non viable for the monopolist (a).
}}
\label{phasediagram.eps}
\end{figure}

In this section we analyse and discuss the solution of the optimal supply-demand static equilibria,
that is, the solutions of equations (\ref{eq.price}) and (\ref{cond2}). 
As might be expected, the result for the product $\beta p_M$ depends 
only on the two parameters $\beta h$ and $\beta J$. That is, the variance 
of the idiosyncratic part of the reservation prices fixes the scale of the 
important parameters, and in particular that of the optimal price.

Let us first discuss the case where $h>0$. It is straightforward to check that in 
this case there is a single solution $\eta_M$. It is interesting to compare the value of $p_M$ 
with the value $p_n$ corresponding to the neutral situation on the demand side. 
The latter corresponds to the {\it unbiased} situation where
the willingness to pay is neutral on average: there are as many agents
likely to buy as not to buy ($\eta=1/2$). Since the expected willingness to pay of any agent $i$ is $h+\theta_i+J/2-p$, its average over the set of agents is $h+J/2-p$. Thus, the neutral
state is obtained for
\begin{equation}
\label{eq.Pn}
p_n=h+J/2.
\end{equation} 
To compare $p_M$ with $p_n$, it is convenient to rewrite equation (\ref{eq.price_d}) as 
\begin{equation}
\beta(p^d-p_n)=\beta J (\eta-1/2) + \ln[(1-\eta)/\eta]. 
\end{equation}
This equation gives $p^d=p_n$ for $\eta=0.5$, as it should. For this value of $\eta$, 
equation (\ref{eq.price_s}) 
gives $p^s=p_n$ only if $\beta (h+J)=2$: for these values 
of $J$ and $h$, the monopolist maximises his profit when the buyers represent 
half of the population. When $\beta (h+J)$ increases above $2$ (decreases below $2$),
 the monopolist's optimal price decreases (increases) and the corresponding 
fraction of buyers increases (decreases).

Finally, if there are no social effects 
($J=0$) the monopolist optimal price is a solution of the implicit equation:  
\begin{equation}
p_M = \frac{1}{\beta \; F(p_M-h)} =\frac{1 + \exp{(- \beta(p_M-h))}}{\beta} .
\end{equation}
The value of $\beta p_M$ lies between 1 and $1+\exp(\beta h)$. Increasing $\beta$ lowers the optimal price: since the variance of the distribution of willingness to pay gets smaller, the only way to keep a sufficient number of buyers is to lower the prices. 

Consider now the case with $h<0$, that is, on average the population 
is not willing to buy. Due to the randomness of the individual's reservation 
price, $H_i=H+\theta_i$, the surplus may be positive but only for a small fraction 
of the population. Thus, we would expect that the monopolist will maximise 
his profit by adjusting the price to the preferences of this minority. However, 
this intuitive conclusion is not supported by the solution to equations (\ref{eq.price}) 
when the social influence represented by $J$ is strong enough. The optimal 
monopolist's strategy shifts abruptly from a regime of high price and a small 
fraction of buyers to a regime of low price with a large fraction of buyers 
as $\beta J$ increases. Such a discontinuity may actually be expected for $\beta J >4=\beta J_B$,
that is when the demand itself has a discontinuity. But, quite interestingly, 
the transition is also found in the range $\beta J_A \equiv 27/8 \; < \beta J < 4==\beta J_B$, that is, in a domain
of the parameters space $(\beta J, \beta h)$ where the demand $\eta(p)$ is a
smooth function of the price. 

Such a transition is analogous to what is called a first order 
phase transition in physics~\cite{Stanley}: the fraction of buyers jumps 
at a critical value of the control parameter $\beta J_c(\beta h)$ from a low to a high value. 
Before the transition, above a  value $\beta J_{-}(\beta h) < \beta J_c(\beta h)$ 
equations (\ref{eq.price}) already present several solutions. Two 
of them are local maxima of the monopolist's profit function, 
and one corresponds to a local minimum. The global maximum is the solution corresponding to a high price 
with few buyers for $\beta J < \beta J_c$, and that of low price with many buyers  for $\beta J > \beta J_c$. 
Figure \ref{fig.supply} presents these results for the particular value $\beta h=-2$, for which it can be shown 
analytically that $\beta J_{-}=4$, and $\beta J_c \approx 4.17$ (determined  numerically).

The detailed discussion of the full phase diagram in the plane $\{\beta J, \beta h\}$, shown on Figure \ref{phasediagram.eps},
is presented in the Appendix, and a more general discussion will be presented elsewhere \cite{extended}.

\section{Dynamic features}
\label{sec.dynamic}

In this class of models, the individual threshold of adoption implicitly
embodies the number of agents each individual considers sufficient to modify his
behaviour, as underlined in the field of social science~\cite{Schelling78,Granovetter78}. 
We briefly discuss here some dynamic aspects, considering a market dynamics
with myopic customers: an agent makes its decision at time $t$ based on the observation
of the behaviour of other agents at time $t-1$. 
The adoption by a single agent in the population (a ``direct adopter'') 
may then lead to a significant change in the whole population through a chain reaction of
``indirect adopters''~\cite{PhanPajotNadal03}. 

\begin{figure} [htbp]
\centering
\begin{tabular}{|c|c|}
\hline
\\
\includegraphics[width=3.6cm]{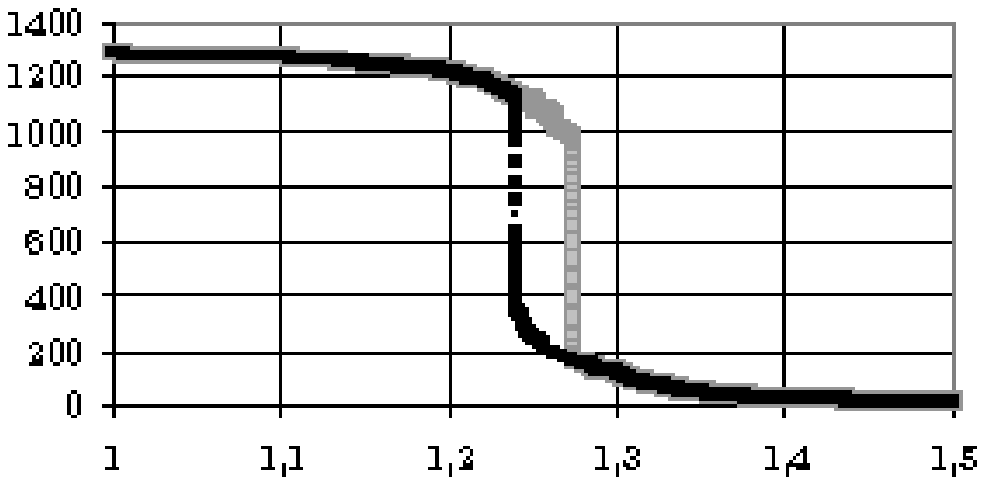} &
\includegraphics[width=3.2cm]{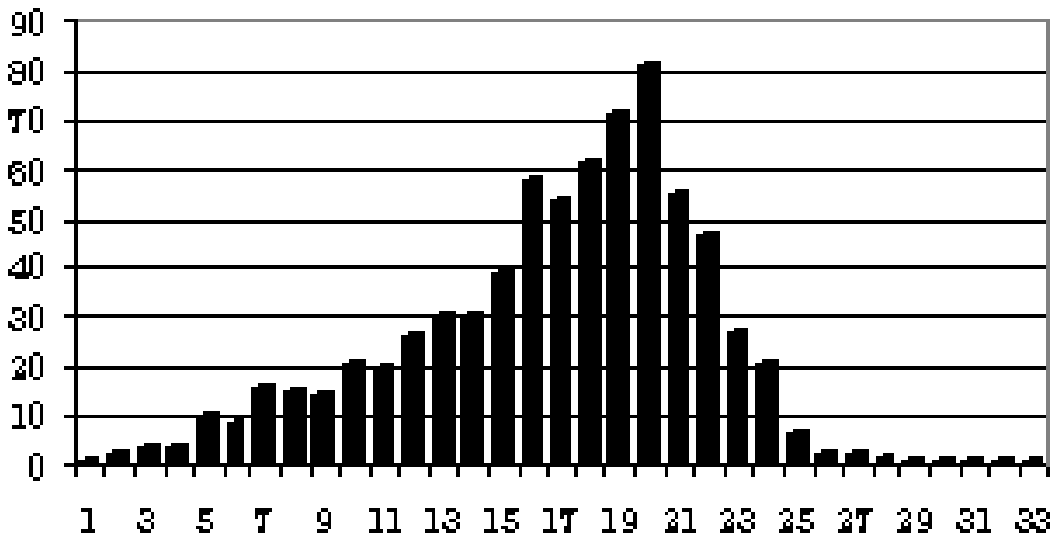} \\
\scriptsize{a - Hysteresis in the trade-off between prices} &
\scriptsize{b - Chronology and sizes of induced effects}  \\
\scriptsize{and customers: upstream (black)} &
\scriptsize{ in an avalanche at the phase} \\
\scriptsize{and downstream (grey) trajectories}&
\scriptsize{transition for $P= 1.2408$ ($P_n=1.25$)}
\\ \hline
\end{tabular}
\\
\caption[]{ \scriptsize{Discontinuous phase transition (full connectivity, synchronous
activation regime; source: Phan {\it et al.}~\cite{PhanPajotNadal03}; 
parameters: $N=1296$, $H=1$, $J=0.5$, $\beta$ = 10).}
}
\label{phaseTransition}
\end{figure}

Within the McF model, the dynamics
for the fraction of adopters in the large $N$ limit is then given by
\begin{equation}
\label{eq.dyn}
\eta(t) = 1 - F(P - H - J \; \eta(t-1))
\end{equation}
and $\eta(t)$ converges to a solution of the fixed point equation (\ref{eq.fp}).
As we have seen, given $J$, $H$ and $P$, two stable and one unstable fixed points appear in (\ref{eq.fp}) 
for values of $\beta$ large enough (small $\sigma$). 
The stable solutions correspond to two possible levels of $\eta$ at a given price
(Figure \ref{phaseTransition}a).
Varying the price smoothly, a transition may be observed between these phases.
The jump in the number of buyers occurs at different price values according
to whether the price increases or decreases,
leading to {\em hysteresis loops}. In some cases, the number of customers
evolves through a series of clustered flips (between $\omega_i=1$ and $\omega_i=0$), called avalanches.
For small values of $\beta$ (large $\sigma$), 
there is a single fixed point for all values of $P$, 
and no hysteresis at all (~\cite{PhanPajotNadal03,PhanNadal03}).

The curves in Figure \ref{phaseTransition}a, represent the number of customers 
as a function of the price, obtained through a simulation of the whole demand 
system. The black (grey) curve corresponds to the ``upstream'' (downstream) trajectory,
when prices decrease (increase) in steps of $10^{-4}$, within the interval
$[0.9, 1.6]$. We observe a {\it hysteresis} phenomenon with discontinuous 
transitions around the theoretical point of symmetry, $P_n = 1.25$. 
Typically, along the downstream trajectory (with increasing prices, grey curve) the
externality  effect induces a strong resistance of the demand system against
a decrease in the number of customers.
In both cases, large avalanches occur at the so-called ``first order phase transition''.
Figure \ref{phaseTransition}b represents the sizes of these dramatic induced effects as a function of time (see \cite{PhanPajotNadal03,PhanNadal03} for more details).

\section{Conclusion}

In this paper, we have compared two extreme special cases of discrete choice models,
the McFadden (McF) and the Thurstone (TP) models, in which the individuals bear
a local positive social influence on their willingness to pay, and have random heterogeneous
idiosyncratic preferences. In the McF model the latter remain fixed, and
give rise to a complex market organisation. For physicists, this model with fixed
heterogeneity belongs to the class of `quenched' disorder models; the McF model is
equivalent to a `Random Field Ising Model' (RFIM). In the TP model, all the agents
share a homogeneous part of willingness to pay, but have an additive, time varying,
random (logistic) idiosyncratic characteristic. In physics, this problem corresponds
to a case of `annealed' disorder. In the TP model, the 
random idiosyncratic component is equivalent to having a stochastic dynamics, because 
each agent decides to buy according to the logit choice function, making this 
model formally equivalent to an Ising model at temperature $T \neq 0$ in a uniform 
(non random) external field. From the physicist's point of view, the McF and TP models 
are quite different: random field and zero temperature in the McF case, uniform field 
and non zero temperature in the TP case. An important result in statistical physics 
is that quenched and annealed disorders can lead to very different behaviours. In  this 
paper we have discussed some consequences on the market's behaviour.

Considering that the seller optimises his own profit, we have exhibited a
new `first order phase transition': when the social influence is strong enough, there
is a regime where, upon increasing the mean willingness to pay, or decreasing the
production costs, the optimal monopolist's solution jumps from one with a high
price and a small number of buyers, to one with a low price and a large number
of buyers.

We have only considered fully connected systems: the theoretical analysis of
systems with finite connectivity is more involved, and requires numerical simulations.
The simplest configuration is one where each customer has only two neighbours, one
on each side. The corresponding network, which has the topology of a ring, has been
analysed numerically by Phan {\it et al.}~\cite{PhanPajotNadal03}) who show that the optimal monopolist's
price increases both with the degree of the connectivity graph and the range of the interactions 
(in particular, in the case of small worlds). Buyers' clusters of different sizes may 
form, so that it is no longer possible to describe the externality with a single 
parameter, like in the mean field case.  Further studies in computational economics 
are required in order to explore such situations.

\pagebreak
\section*{Appendix: Phase Diagram}

In this Appendix we detail the derivation of the phase diagram
in the plane ($\beta J$, $\beta h$), presented in figure \ref{phasediagram.eps}.
The phase diagram shows the domain in the parameter space where
coexist two maxima of the monopolist's profit, one global maximum
(the optimal solution) and one local maximum. Inside this domain there
is a (first order) transition line where, as $\beta J$ and/or $\beta h$
increases, the optimal solution jumps from a solution '-' with a low value
$\eta=\eta_-$ to a solution '+' with a large value $\eta=\eta_+$, $\eta$ being 
the fraction of buyers.
In figure \ref{phasediagram.eps}, the circles 'o' are points on the transition line
obtained numerically; all the other curves being obtained analytically
as explained below.

In the following without loss of generality we set $\beta = 1$, which is equivalent to say that we 
measure $J$ and $h$ in units of $1/\beta$.

To explain the phase diagram in more detail, it is convenient to parametrise every quantity/curve as a function of $\eta$.
First the ({\it per unit}) profit $p$ is given by
\begin{equation}
p= \frac{1}{1-\eta} - J \eta
\label{p(eta)}
\end{equation}
(hence the profit $N \Pi =N p(\eta) \eta $),
and $\eta $ is a fixed point of the equation
\begin{equation}
\eta = G(h,J,\eta )  
\label{(1a)}
\end{equation}
with
\begin{equation}
G(h,J,\eta) = \frac{1}{1+\exp(- h - 2 J \eta + \frac{1}{1-\eta}) }   
\label{(1b)}
\end{equation}

We will also make use of an alternative form of
(\ref{(1a)}), (\ref{(1b)}), that is
\begin{equation}
\label{(1c)}
h = - 2 J \eta   + \frac{1}{1-\eta}  + \log(\frac{\eta}{1-\eta})  
\end{equation}

One can also show that the fixed point equation (\ref{(1a)}) is equivalent to
\begin{equation}
\frac{d(\eta P^d(\eta))}{d\eta} = 0,
\label{equil}
\end{equation}
and that the condition for having a maximum, (\ref{d2pi_L}), is equivalent to
\begin{equation}
\frac{d(\eta P^s(\eta))}{d\eta} > 0.
\label{stab}
\end{equation}

As we will see, there are two singular points of interest:

\begin{tabular}{l l l}
$A$: & $J_A = 27/8$, & $h_A = -3/4 - \log(2)$;
\\
$B$: & $J_B = 4$,  &  $h_B = -2$.
\label{pointsAB}
\end{tabular}

Let us describe the phase diagram considering that,
at fixed $J$, one increases $h$ starting from
some low (strongly negative) value.

If $J < 27/8=J_A$, the optimal solution changes continuously, the fraction
of buyers increasing with no discontinuity from a low to a high value
as $h$ increases.
More generally, outside the grey domain 
in figure \ref{phasediagram.eps} there is a unique solution of the optimisation of the profit.

\begin{figure} 
\vspace{-7.0em}
\centering
\caption[]{\scriptsize{Functions $y=G(h,J,\eta), y=\eta$, price $p(\eta)$ and profit $\Pi(\eta)$ ($+-$).
The intersects of $y=G(h,J,\eta)$ with $y=\eta$ give the extrema of the profit; 
the (possibly local) maxima are those for which $d\Pi/d\eta > 0$. Shown here 
are marginal cases where for one solution $d\Pi/d\eta = 0$, that is  
$y=G(h,J,\eta)$ is tangent to $y=\eta$ (points on the lower or upper curves
of the phase diagram, figure \ref{phasediagram.eps}).}}
\begin{tabular}{c c}
\includegraphics[height=3cm]{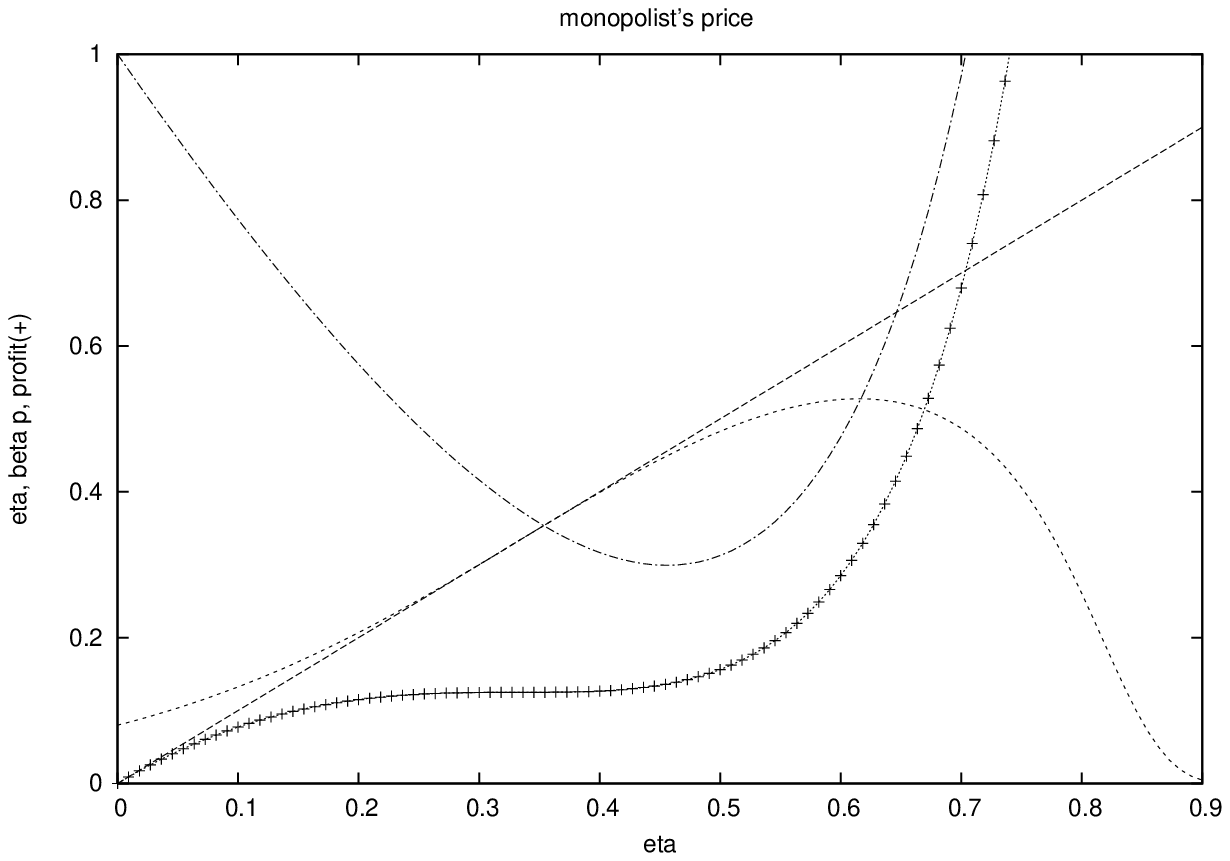} &
\includegraphics[height=3cm]{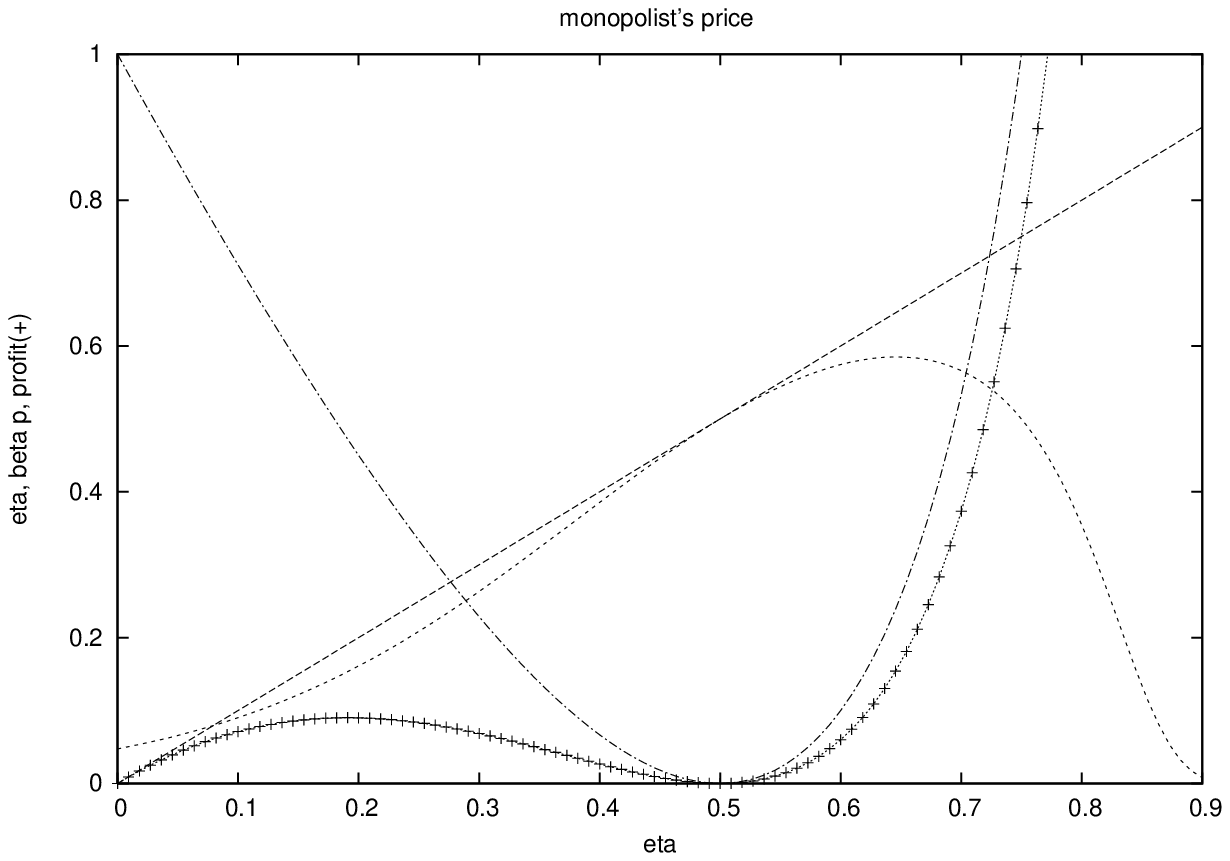} \\
\scriptsize{(a) $J=27/8$ $h= -3/4 - \log(2)$} &
\scriptsize{(b) $J=4$ $h=-2$} \\ 
\includegraphics[height=3cm]{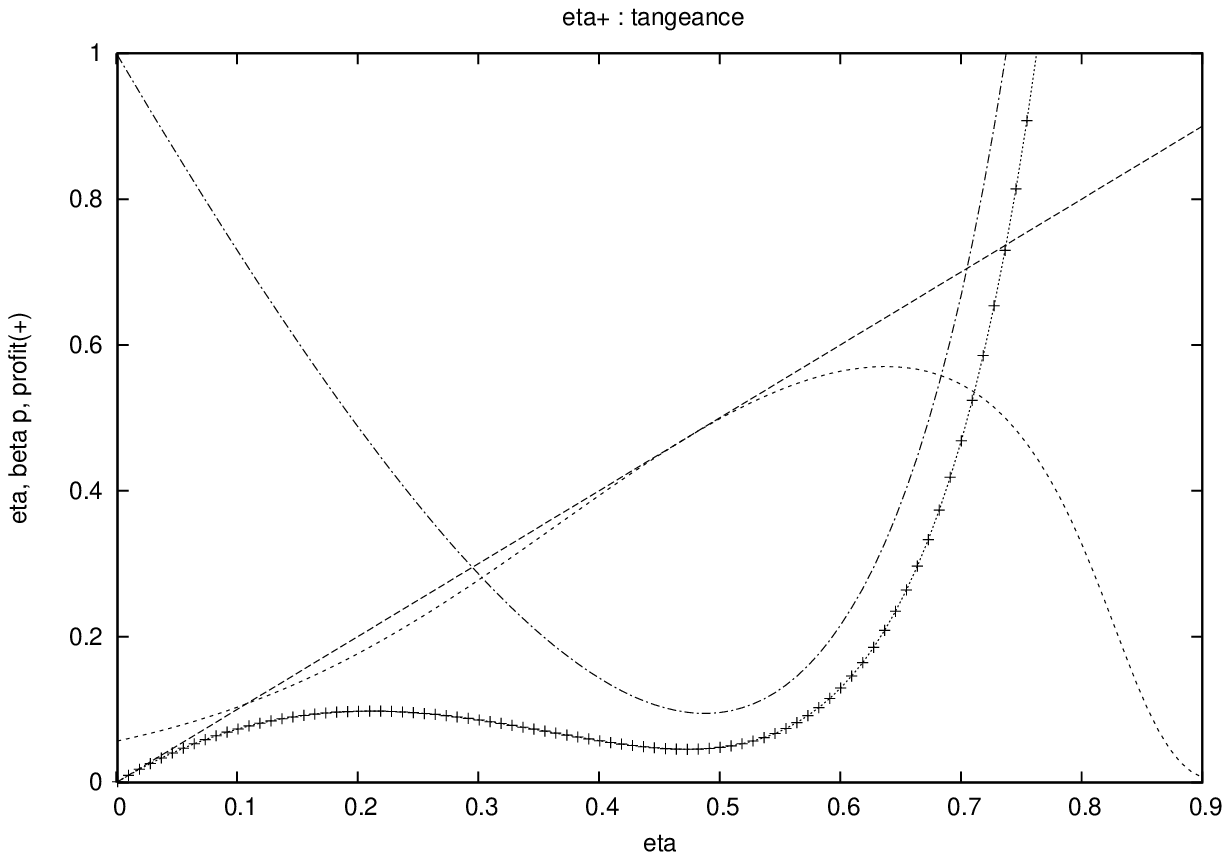} &
\includegraphics[height=3cm]{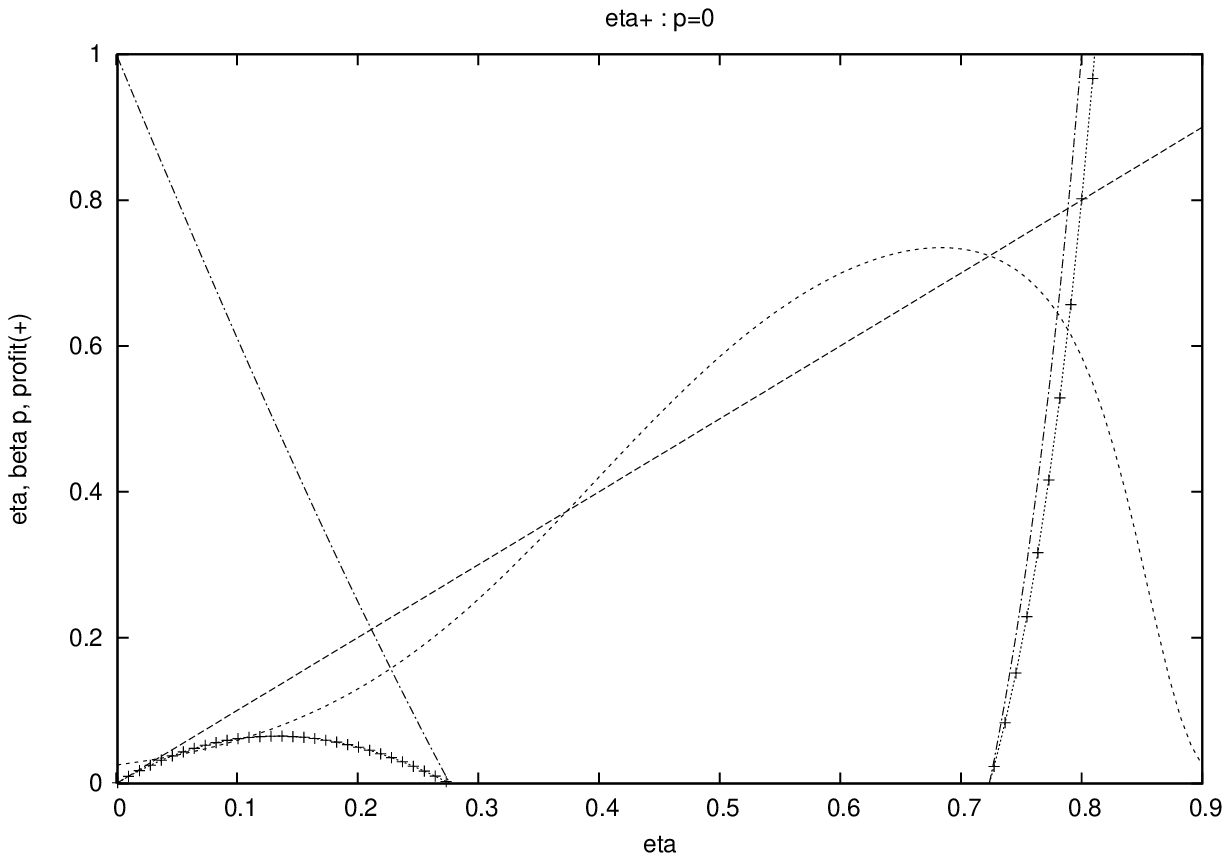} \\
\scriptsize{(c1) Point on the lower curve, $\eta < 1/2$} &
\scriptsize{(c2) Point on the lower curve, $\eta > 1/2$} \\
\scriptsize{(d1) Point on the upper curve, $J > 4$} & 
\scriptsize{(d2) Point on the upper curve, $J < 4$} \\
\includegraphics[height=3cm]{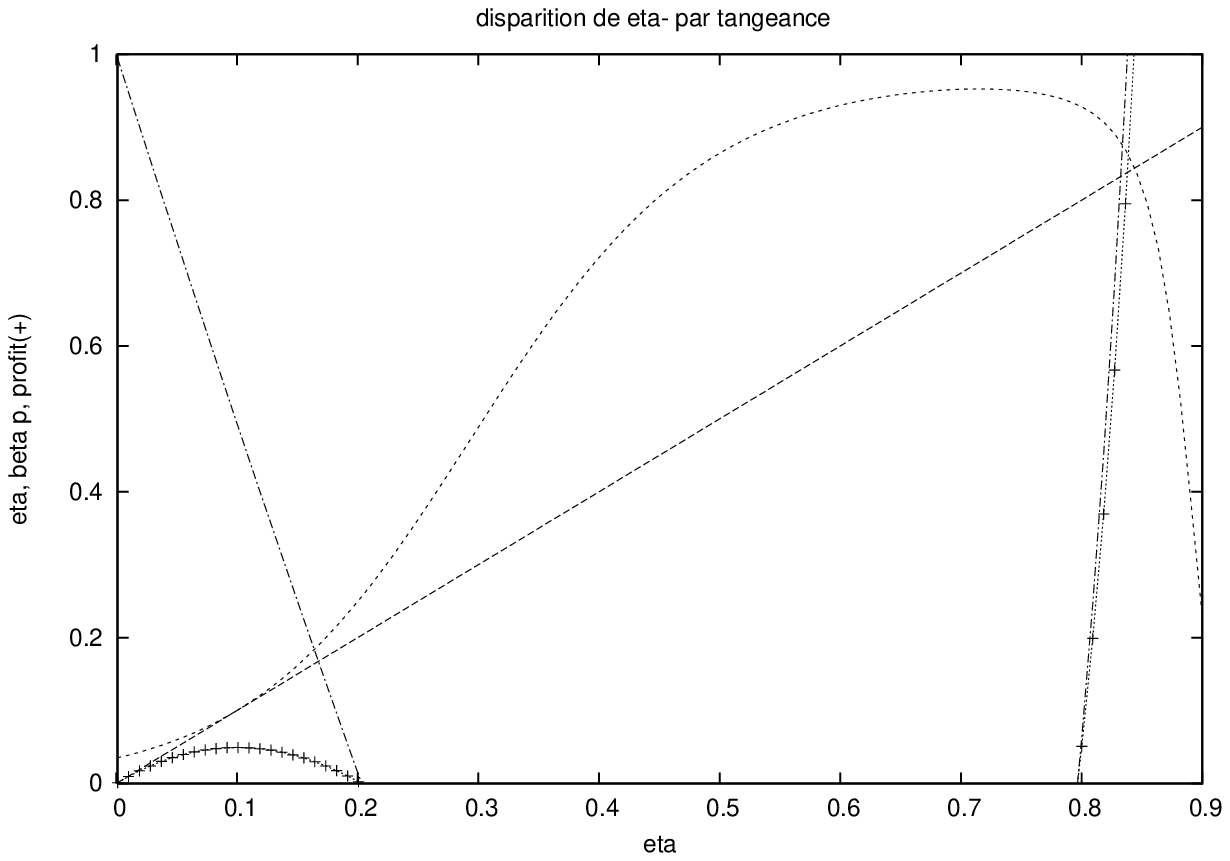} &
\includegraphics[height=3cm]{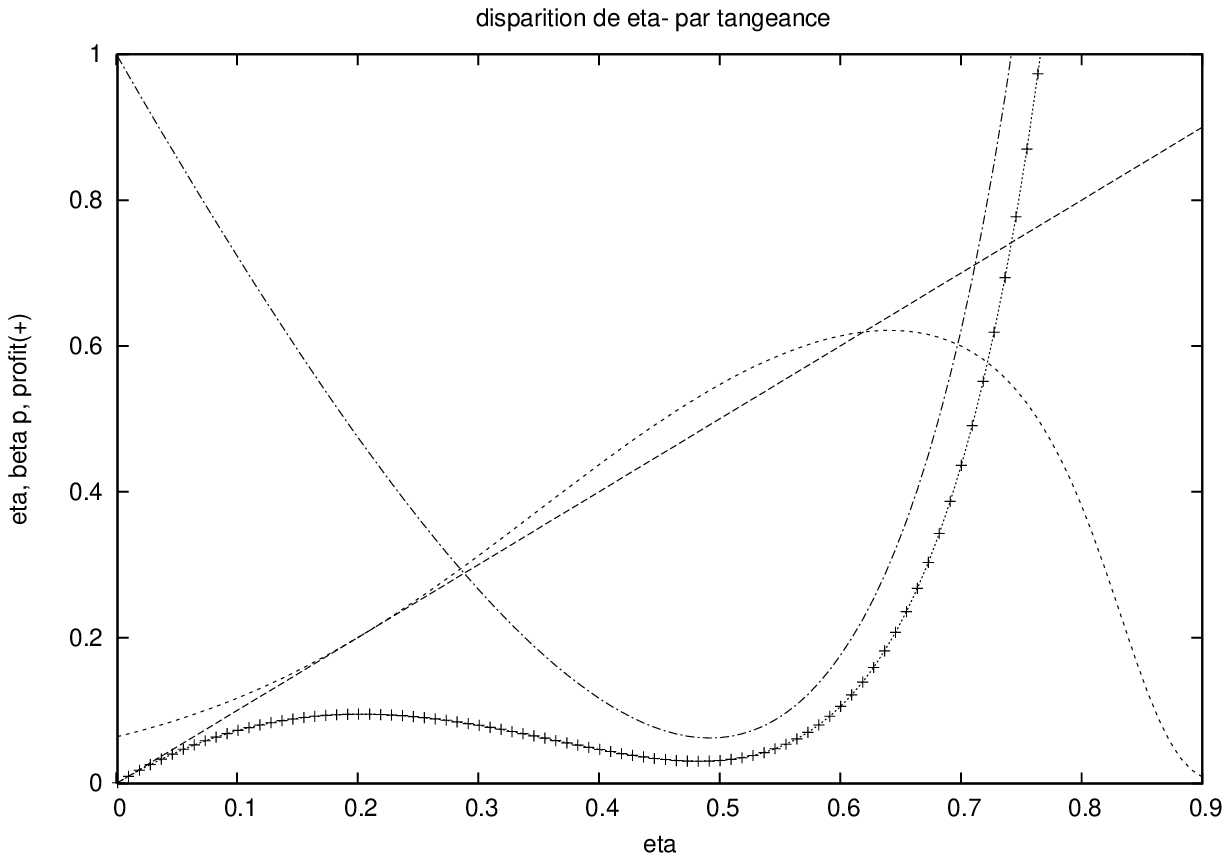} \\
\end{tabular}
\label{smallfigures}
\end{figure}

For $J > 27/8=J_A$, as $h$ increases one will first hit the lower boundary of the grey region on the phase 
diagram, $h=h_{-}(J)$. On this line, a local maximum of the profit
appears, corresponding to a value $\eta=\eta_+ > 1/3$.
As shown in figure \ref{smallfigures}a 
, the curve $y= G(h,J,\eta)$
intersects $y=\eta$ at some small value $\eta=\eta_{-}$ and
is tangent to it at $\eta=\eta_+$. For $h_{-}(J) < h < h_{+}(J)$
$y= G(h,J,\eta)$ has three intersects with the diagonal $y=\eta$,
$h = h_{+}(J)$ being the upper boundary of the grey region on the phase diagram.
The stability analysis shows that the two extreme intersects correspond
to maxima of the profit, giving the solutions $\eta=\eta_{-}$ and $\eta=\eta_+$.
On the upper boundary $h = h_{+}(J)$, it is the solution with a small value of $\eta$
which disappears, 
with $y= G(h,J,\eta)$ becoming tangent to $y=\eta$ for $\eta=\eta_{-}$,
see figures \ref{smallfigures}c1 and \ref{smallfigures}c2. 
These lower and upper boundaries are obtained by writing that
the second derivative of the profit with respect to $p$ is zero, giving 
\begin{equation}
\label{(2)}
2 J \eta (1-\eta)^2 = 1   
\end{equation}
Together with (\ref{(1c)}) 
this gives the curves parametrised by $\eta$,
\begin{eqnarray}
\label{(3)}
J & = & \frac{1}{ 2 \eta (1-\eta)^2} \nonumber \\
h & = & - \frac{1}{(1-\eta)^2}  + \frac{1}{1-\eta}  + \log(\frac{\eta}{1-\eta})
\end{eqnarray}
the lower curve $h_{-}(J)$ corresponding to the branch $\eta = \eta_{+} \in [1/3,\; 1]$, and
the upper curve $h_{+}(J)$ corresponding to the branch $\eta = \eta_{-} \in [0, \; 1/3]$.

The two curves merge at the singular point $A$, at which $\eta_{+} = \eta_{-} = 1/3$,
$J_A=27/8, h_A=-3/4 - \log(2)$.
Expanding the above equations (\ref{(3)}) near $\eta= 1/3$ we find
that the two curves are cotangent at $A$, with a slope $-2/3$.
This common tangent is thus also tangent to the transition line at $A$.
A straight segment of slope $-2/3$ starting from $A$ is plotted on the
phase diagram, figure \ref{phasediagram.eps}, and one can see that this is a very good approximation
of the transition line for $J < 4=J_B$.

On the lower boundary, for $J > 4=J_B$, the local maximum with $\eta=\eta_{+}$
appears with a negative profit (zero profit at point $B$ where $\eta_{+} = 1/2$). 
The profit becomes positive
on the curve starting at point $B$, on which the profit is zero with $\eta_{+} > 1/2$.
This curve is obtained by writing $p = 0$, $\eta_{+} > 1/2$, that is,
\begin{equation}
\eta_{+} = \frac{1}{2} \left[1+ \sqrt{1- \frac{4}{J}}\right]
\label{eta_p=0}
\end{equation}
and $h$ is obtained as a function of $J$, by replacing $\eta$ in (\ref{(1c)}) 
by the above expression (\ref{eta_p=0}).
In this domain of negative profit for the local maximum, the distance to the transition line
(at a given value of $J$) is equal to the amount by which the production cost
per unit of good must be lowered in order to make the solution viable.

In the domain $J > 4=J_B$, the transition line, computed numerically, appears to be 
just above this null profit line.
This suggests that an expansion for $p$ small for the
solution $\eta_{+}$, and for $\eta$ small for the solution
$\eta_{-}$ should provide good approximations.
The transition is obtained when $\Pi_{+} = \Pi_{-}$ as explained below,
and this allows to display the curve of figure \ref{phasediagram.eps},
which turns out to be a very good approximation of the transition line for large values of $J$
(or small values of $h$, typically $h < - 4.5$).

Let us first consider the vicinity of the point $B$
at which $p_{+}=0$, $\eta_{+}=1/2$, and the second derivative of the profit
is zero for this '+' local solution.
Expanding near $J=4$, $h$ just above $h_{-}(4)=-2$ ($p$ small),
one gets the behaviour of the '+' solution:
\begin{eqnarray}
\label{BJ4}
\epsilon & \equiv & h + 2, \;\; 0 < \epsilon << 1 \nonumber \\
\eta_{+} & = & \frac{1}{2} ( 1 + \sqrt{\frac{\epsilon}{2}} )\nonumber \\
p_{+}& = & \epsilon +  o(\epsilon^{3/2}) \nonumber \\
\Pi_{+} & = & \frac{\epsilon}{2} + o(\epsilon^{3/2})
\end{eqnarray}
The singular, square-root, behaviour of $\eta$ is specific to point $B$.
For any $J>4$, just above the null curve $\eta_{+}$ increases linearly with
$\epsilon \equiv h-h_{-}(J)$, 
but with a similar behaviour as for $J=4$
for the price and the profit,
at lowest order in $\epsilon$: denoting by $\eta_{+}^0(J)$
the value $\eta_{+}(h_{-}(J))$,
\begin{eqnarray}
0 & < & \epsilon=h-h_{-}(J) << 1 \nonumber \\
\eta_{+} & = & \eta_{+}^0(J) + \epsilon \frac{\eta_{+}^0 (1-\eta_{+}^0)^2}{1 - 2 J \eta_{+}^0 (1-\eta_{+}^0)^2} \nonumber \\
p_{+}& = & \epsilon, \nonumber \\
\Pi_{+} & = & \eta_{+}^0(J) \; \epsilon.
\label{etapPi-}
\end{eqnarray}
One can see from the expression of $\eta_{+}$ in (\ref{etapPi-}) how the singularity at point $B$
appears: the coefficient of $\epsilon$ diverges when condition (\ref{(2)}) is fulfilled, 
that is when the solution is marginally stable, which
is the case at $B$.

Similarly one can get the behaviour of the '+' solution
near point $B$ at $h=-2$,
increasing $J$ from $J=4$, as shown
in figure \ref{fig.supply} 
\begin{eqnarray}
\label{Bh-2}
\eta_{+} & = & \frac{1}{2} ( 1 + \sqrt{\frac{J-4}{2}} )  \nonumber \\
p_{+}& = & \frac{1}{2} (J-4) + \frac{\sqrt{2}}{12} (J-4)^{3/2}  \nonumber \\
\Pi_{+} & = & \frac{1}{4} (J-4) + \frac{1}{3\sqrt{2}}  (J-4)^{3/2} 
\end{eqnarray}

Coming back to the behaviour at a given value of $J$,
one can get an approximation of the '-' solution. The
 fixed point equation for $\eta$
for given values of $J$ and $h$, is
\begin{equation}
\eta = H(\eta) \equiv 1 / \left[1 + \exp{(1-h- 2 J \eta + \frac{\eta}{1-\eta})} \right]
\label{fp_eta}
\end{equation}
The '-' solution corresponding to a small value
of $\eta$ can be found by iterating $\eta(k+1)=H(\eta(k))$
starting with $\eta(0)=0$, and $\eta(k)$ is an increasing
sequence of approximations of $\eta_{-}$. The 
lowest non trivial order is then given by
\begin{equation}
\eta_{-}^0 = H(0) = 1 / \left[1 + \exp{(1-h)} \right]
\label{fp0}
\end{equation}
which is indeed small for $h$ strongly negative. At the next order
\begin{equation}
\eta_{-}^1 = H(\eta_{-}^0)
\label{fp1}
\end{equation}
Taking $\eta_{-}^0$ as the small parameter, the expansion of $\eta_{-}^1$
gives 
\begin{equation}
\eta_{-}^1 = \eta_{-}^0 ( 1 + (2 J -1) \eta_{-}^0)
\label{eta1}
\end{equation}
and this gives the corresponding approximations
for the price and the profit,
\begin{eqnarray}
p_{-}^1 & = & 1 - (J-1) \eta_{-}^0 \nonumber \\
\Pi_{-}^1 & = & \eta_{-}^0  + J (\eta_{-}^0)^2.
\label{pr-1}
\end{eqnarray}

It is clear from the above equation that the dependency on $J$ is weak
since $\eta_{-}^0$ is small, in agreement with the exact behaviour computed
numerically, shown on figure \ref{fig.supply}.

Now we consider the neighbourhood of $(J, h_{-}(J))$, that is
$h= h_{-}(J) + \epsilon$. Denoting by
$\eta_0(J)$ the value of $\eta_{-}^0$ at $h=h_{-}(J)$,
$\eta_{-}^0 (h) = \eta_0(J)  +  \epsilon \eta_0(J) (1 - \eta_0(J))$.
Taking this expression for computing the profit of the '-' solution,
and writing
that at the transition the two solutions '+' and '-' give the same profit,
one gets the following approximation for the value $\epsilon_c(J)$
of $\epsilon$
at the transition
(hence the value of $h$ at the transition, $h_c(J)= h_{-}(J) + \epsilon_c(J)$):
\begin{equation}
\epsilon_c (J)=  \frac{\eta_0(J)}{\eta_{+}(J)}
\label{hc}
\end{equation}
where $\eta_{+}(J)$ is given by equation (\ref{etapPi-}).
It is this  curve $h_c(J)= h_{-}(J) + \epsilon_c (J)$ which is plotted in
figure \ref{phasediagram.eps}  for $J > J_B=4$.
 
\clearpage
\clearpage

\end{document}